\documentclass[a4paper,11pt]{article}
\usepackage{pos}

\usepackage{graphicx}
\usepackage{dcolumn}
\usepackage{xcolor}
\usepackage{hyperref}

\title{Disconnected contribution to the LO HVP term of muon g-2 from ETMC
}

\author[a,b]{C.~Alexandrou}
\author*[b]{S.~Bacchio}
\author[c]{P.~Dimopoulos}
\author[b]{J.~Finkenrath}
\author[d]{R.~Frezzotti}
\author[e]{G.~Gagliardi}
\author[f]{M.~Garofalo}
\author[a,b]{K.~Hadjiyiannakou}
\author[g]{B.~Kostrzewa}
\author[h]{K.~Jansen}
\author[i]{V.~Lubicz}
\author[f]{M.~Petschlies}
\author[e]{F.~Sanfilippo}
\author[e]{S.~Simula}
\author[f]{C.~Urbach}
\author[j]{U.~Wenger}

\affiliation[a]{Department of Physics, University of Cyprus, 20537 Nicosia, Cyprus}
\affiliation[b]{Computation-based Science and Technology Research Center, The Cyprus Institute,\\20 Konstantinou Kavafi Street, 2121 Nicosia, Cyprus}
\affiliation[c]{Dipartimento  di  Scienze  Matematiche,  Fisiche  e  Informatiche,  Universit\`a  di  Parma  and  INFN, Gruppo  Collegato  di  Parma,  Parco  Area  delle  Scienze  7/a  (Campus),  43124  Parma,  Italy}
\affiliation[d]{Dipartimento di Fisica and INFN, Universit\`a di Roma ``Tor Vergata",\\Via della Ricerca Scientifica 1, I-00133 Roma, Italy}
\affiliation[j]{Institute for Theoretical Physics, Albert Einstein Center for Fundamental Physics,\\University of Bern, Sidlerstrasse 5, CH-3012 Bern, Switzerland}
\affiliation[e]{Istituto Nazionale di Fisica Nucleare, Sezione di Roma Tre,\\Via della Vasca Navale 84, I-00146 Rome, Italy}
\affiliation[f]{HISKP (Theory), Rheinische Friedrich-Wilhelms-Universit\"at Bonn,\\Nussallee 14-16, 53115 Bonn, Germany}
\affiliation[g]{High Performance Computing and Analytics Lab, Rheinische Friedrich-Wilhelms-Universit\"at Bonn,\\ Friedrich-Hirzebruch-Allee 8, 53115 Bonn, Germany}
\affiliation[h]{NIC, DESY, Platanenallee 6, D-15738 Zeuthen, Germany}
\affiliation[i]{Dipartimento di Matematica e Fisica, Universit\`a Roma Tre and INFN, Sezione di Roma Tre,\\Via della Vasca Navale 84, I-00146 Rome, Italy}

\emailAdd{s.bacchio@gmail.com}

\abstract{We present a lattice determination of the disconnected contributions to the leading-order hadronic
vacuum polarization (HVP) to the muon anomalous magnetic moment in the so-called short and intermediate time-distance windows. We employ gauge ensembles produced by the Extended Twisted
Mass Collaboration (ETMC) with $N_f = 2 + 1 + 1$ flavours of Wilson twisted-mass clover-improved
quarks with masses approximately tuned to their physical value. We take the continuum limit employing three lattice spacings at about 0.08, 0.07 and 0.06 fm.}

\FullConference{%
  The 39th International Symposium on Lattice Field Theory (Lattice2022),\\
  8-13 August, 2022 \\
  Bonn, Germany 
}

\begin{document}
\maketitle

\section{Introduction}

In Ref.~\cite{Alexandrou:2022amy} we have presented our full calculation of the short and intermediate time-distance hadronic vacuum polarization (HVP) contributions to the muon magnetic moment using twisted-mass fermions
on ETMC ensembles. In this work, we present an extract that focuses on disconnected contributions, summarizing the results of the manuscript and providing some additional details.

\begin{figure}[htb!]
    \centering
    \hfill
    \includegraphics[width=0.4\textwidth]{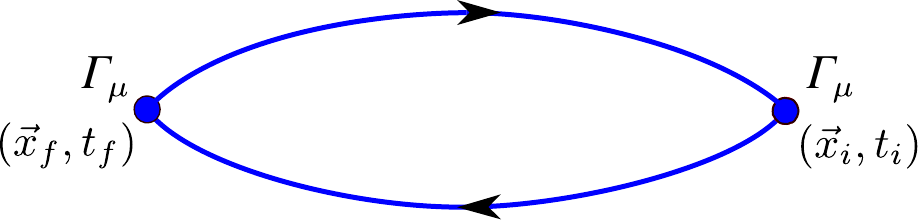}\hfill
    \includegraphics[width=0.4\textwidth]{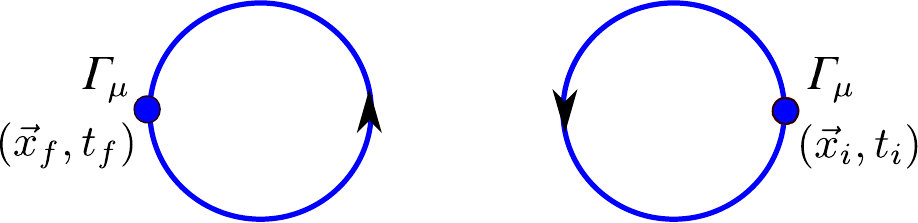}\hfill
    \caption{\it \small Connected (left) and disconnected (right) contributions to the vector correlator $V(t)$.}
    \label{fig:diagrams}
\end{figure}

In our calculation, we have adopted the time momentum representation\,\cite{Bernecker:2011gh} and evaluate the HVP contribution to the muon anomalous magnetic moment $a_{\mu}^{\rm HVP}$ as
\begin{equation}
    \label{eq:amu_HVP}
    a_{\mu}^{\rm{HVP}} = 2 \alpha_{em}^2 \int_0^\infty ~ dt \, t^2 \, K(m_\mu t) \,V(t) ~ , ~  
\end{equation}
where $t$ is the Euclidean time and the kernel function $K(m_{\mu} t)$ is defined as\footnote{The leptonic kernel $K(z)$ is proportional to $z^2$ at small values of $z$ and it goes to $1$ for $z \to \infty$.}
\begin{equation}
    \label{eq:kernel}
    K(z) = 2 \int_0^1 dy ( 1- y) \left[ 1 - j_0^2 \left(\frac{z}{2}\frac{y}{\sqrt{1 - y}} \right) \right]~,\qquad j_{0}(y) = \frac{\sin{(y)}}{y} ~ . ~
\end{equation}
The Euclidean vector correlator $V(t)$ is defined as
\begin{equation}
     \label{eq:VV}
     V(t) \equiv - \frac{1}{3} \sum_{i=1,2,3} \int d^3{x} ~ \langle J_i(\vec{x}_f, t_f) J_i(\vec{x}_i, t_i) \rangle 
\end{equation}
with $J_\mu(x)$ being the electromagnetic current operator
\begin{equation}
      \label{eq:Jmu}
     J_\mu(x) \equiv \sum_{f = u, d, s, c, ...} q_f ~ \overline{\psi}_f(x) \gamma_\mu \psi_f(x) ~ 
\end{equation}
 and $q_f$ the electric charge for the quark flavour $f$ (in units of the absolute value of the electron charge). 
Clearly, the vector correlator $V(t)$ give rise to both connected and disconnected contributions, as depicted in Fig.~\ref{fig:diagrams}. The latter are the focus of this proceeding.

\subsection{The RBC/UKQCD windows in the time-momentum representation}
\label{sec:windows_t}

Following the analysis of the RBC/UKQCD Collaboration\,\cite{RBC:2018dos}, we separate the whole HVP contribution in three terms, 
\begin{equation}
\label{eq:amuw}
    a_\mu^{\rm HVP} \equiv a_\mu^{\rm SD} + a_\mu^{\rm W} + a_\mu^{\rm LD} ~ , ~
\end{equation}
that can be obtained from  Eq.\,(\ref{eq:amu_HVP})  with integration kernel $K(m_\mu t)$ multiplied by suitably smoothed Heaviside step-functions, namely
\begin{equation}
    \label{eq:amu_w}
    a_\mu^w = 2 \alpha_{em}^2 \int_0^\infty ~ dt \, t^2 \, K(m_\mu t) \, \Theta^w(t)\,V(t) ~ \qquad w = \{\rm SD, W, LD \} ~ , ~ 
\end{equation}
where the time-modulating function $\Theta^w(t)$ is given by
\begin{eqnarray}
      \label{eq:Mt_SD}
      \Theta^{\rm SD}(t) & \equiv & 1 -  \frac{1}{1 + e^{- 2 (t - t_0) / \Delta}} ~ , ~ \\[2mm]
      \label{eq:Mt_W}
      \Theta^{\rm W}(t) & \equiv & \frac{1}{1 + e^{- 2 (t - t_0) / \Delta}} -  \frac{1}{1 + e^{- 2 (t - t_1) / \Delta}} ~ , ~ \\[2mm]      
      \label{eq:Mt_LD}
      \Theta^{\rm LD}(t) & \equiv & \frac{1}{1 + e^{- 2 (t - t_1) / \Delta}} ~  
\end{eqnarray}
with the parameters $t_0, t_1, \Delta$ chosen\,\cite{RBC:2018dos} to be equal to
\begin{equation}
    \label{eq:parms}
    t_0 = 0.4 ~ \rm{fm} ~ , ~ \qquad t_1 = 1 ~ {\rm fm} ~ , ~ \qquad \Delta = 0.15~{\rm fm} ~ . ~    
\end{equation}

\section{Lattice setup}
In this work we compute disconnected contributions on three ensembles with $N_f = 2 + 1 + 1$ flavours of Wilson twisted-mass clover-improved
quarks with masses approximately tuned to their physical value.
The parameters of the ensembles are give in Table~\ref{tab:simudetails}. For the full calculation in Ref.~\cite{Alexandrou:2022amy}, a fourth ensemble, cB211.072.96, at $\beta=1.778$ and lattice volume $V/a^{4}=96^{3}\cdot 192$ was used for estimating finite size effects (FSEs). However, due to the high cost of the calculation, we did not compute disconnected contributions using the larger volume cB211.072.96 ensemble, since FSEs are expected to be within statistical errors.

\begin{table}[htb!]
\begin{center}
\small
    \begin{tabular}{||c||c|c|c|c|c||c|c||}
    \hline
    ~~~ ensemble ~~~ & ~~~ $\beta$ ~~~ & ~~~ $V/a^{4}$ ~~~ & ~~~ $a$ (fm) ~~~ & ~~~ $a\mu_{\ell}$ ~~~ & ~ $M_{\pi}$ (MeV) ~ & ~ $L$ (fm) ~ & $M_{\pi}L$ ~ \\
  \hline
  cB211.072.64 & $1.778$ & $64^{3}\cdot 128$ & $0.07961~(13)$ & $0.00072$ & $140.2~(0.2)$ & $5.09$ & $3.62$ \\
  
  cC211.060.80 & $1.836$ & $80^{3}\cdot 160$ & $0.06821~(12)$ & $0.00060$ & $136.7~(0.2)$ & $5.46$ & $3.78$ \\
  
  cD211.054.96 & $1.900$ & $96^{3}\cdot 192$ & $0.05692~(10)$ & $0.00054$ & $140.8~(0.2)$ & $5.46$ & $3.90$ \\
  \hline
    \end{tabular}
\end{center}
\caption{\it \small Parameters of the ETMC ensembles used in this work. We give the light-quark bare mass, $a \mu_\ell = a \mu_u = a \mu_d$,  the pion mass $M_\pi$, of the lattice size $L$ and the product $M_\pi L$.}
\label{tab:simudetails}
\end{table} 

\section{Noise reduction techniques and statistics}
The disconnected contributions are computed for the light-, strange- and charm-quark masses. Various noise-reduction techniques are employed to improve the signal-to-noise ratio of disconnected loops. These are the one-end-trick~\cite{McNeile:2006bz}, exact deflation of low-modes~\cite{Gambhir:2016uwp} and hierarchical probing~\cite{Stathopoulos:2013aci}. The one-end-trick is used for all loops; hierarchical probing with distance 8 is used for all loops, except the charm-quark loops for the cB211.072.64 ensemble, where distance 4 is used; and deflation of the low-modes is used for the light quark loops for the cB211.072.64 and cC211.060.80 ensembles. The latter is not  employed for the cD211.054.96 ensemble because of the prohibitively large memory requirements. Indeed, the number of low-modes to be deflated should be increased with the volume, making the costs of this technique to scale with volume-squared. For this reason deflation is not used on the larger volume and, instead, multiple stochastic sources are utilized. In Table~\ref{tab:statistics} we summarize the statistics used for the three ensembles and three quark flavours.

\begin{table}[htb!]
    \centering
    {\small
    \begin{tabular}{||c||c|c|c|c||c|c|c|c||c|c|c|c||}
        \cline{2-13}
        \multicolumn{1}{c|}{} & \multicolumn{4}{c||}{\textbf{cB211.072.64}} & \multicolumn{4}{c||}{\textbf{cC211.060.80}} & \multicolumn{4}{c||}{\textbf{cD211.054.96}}\\
        \hline
        Flavour & $N_{\rm defl}$ & $N_{\rm r}$ & $N_{\rm Had}$ & $N_{\rm vect}$& $N_{\rm defl}$ & $N_{\rm r}$ & $N_{\rm Had}$ & $N_{\rm vect}$& $N_{\rm defl}$ & $N_{\rm r}$ & $N_{\rm Had}$ & $N_{\rm vect}$ \\
        \hline
        Light & 200 & 1 & 512 & 6144 & 450 & 1 & 512 & 6144 & 0 & 8 & 512 & 49152 \\
        Strange & 0 & 2 & 512 & 12288 & 0 & 4 & 512 & 24576 & 0 & 4 & 512 & 24576 \\
        Charm & 0 & 12 & 32 & 4608 & 0 & 1 & 512 & 6144 & 0 & 1 & 512 & 6144 \\
        \hline
        \hline
        \multicolumn{1}{||c||}{$N_{\rm confs}$} & \multicolumn{4}{c||}{$\times$750 configurations} & \multicolumn{4}{c||}{$\times$400 configurations} & \multicolumn{4}{c||}{$\times$500 configurations}\\
        \hline
    \end{tabular}
    }
    \caption{\it \small Noise reduction techniques and statistics used for the disconnected quark loops. For each ensemble, the columns are in order: i) the number of deflated eigenvectors $N_{\rm defl}$, ii) the number of stochastic sources $N_{\rm r}$, iii) the number of Hadamard vectors $N_{\rm Had}$, and iv) the total number of computed vectors $N_{\rm vect} = 12\times N_{\rm r}\times N_{\rm Had}$.}
    \label{tab:statistics}
\end{table}
\begin{figure}[htb!]
    \centering
    \includegraphics[width=\linewidth]{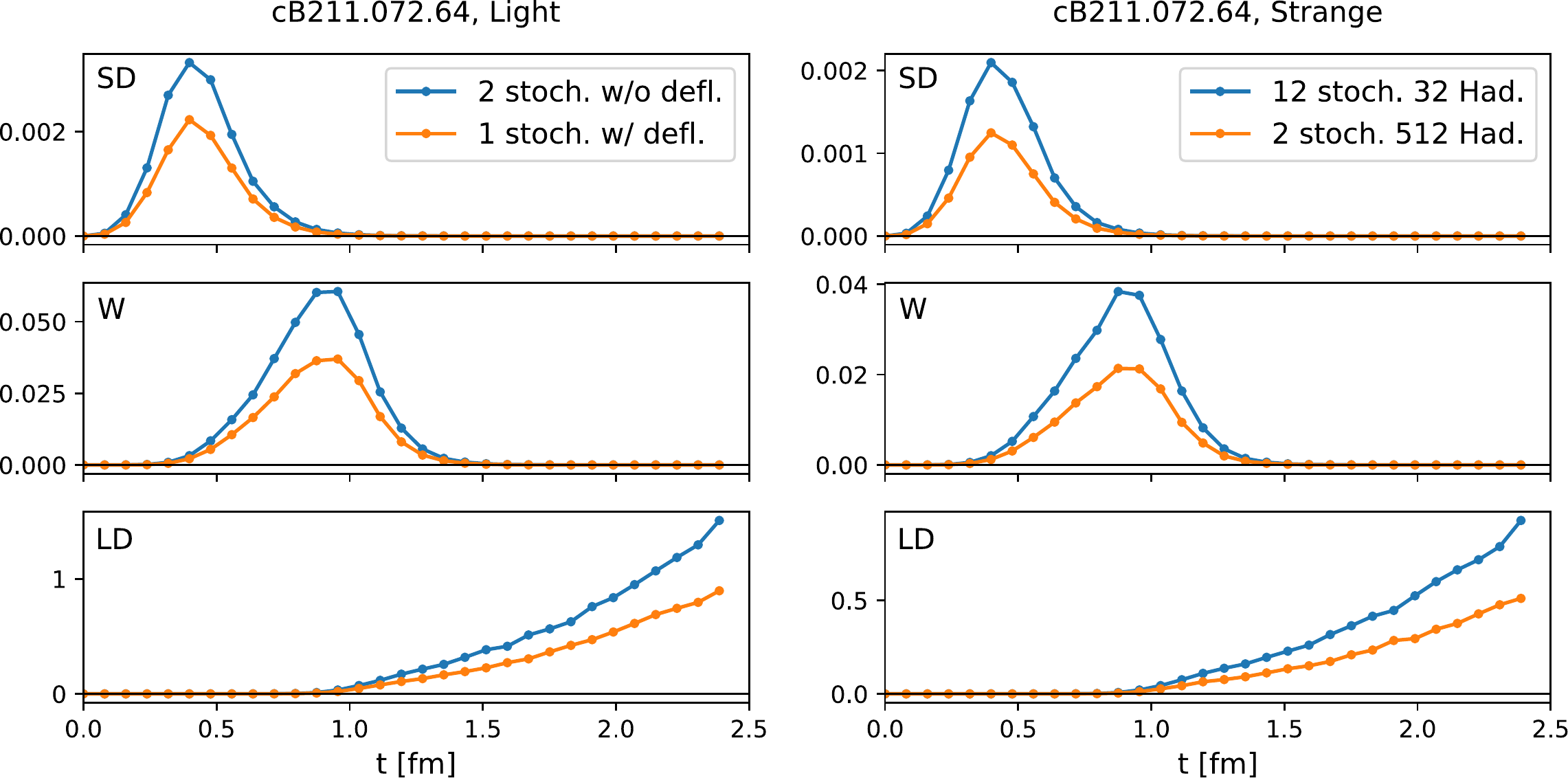}
    \caption{\it \small Statistical errors of the integrand in the short (SD), intermediate (W) and long (LD) distance windows for the light (left panel) and strange (right panel) contributions to the disconnected loops on the cB211.072.64 ensemble. In orange are the values used in the final calculation, i.e. those reported in Table~\ref{tab:statistics}, while in blue are the same quantities computed with an alternative setup. Namely, we used 2 stochastic sources (instead of 1) but no deflation for the light loops and used 12 stochastic sources (instead of 2) and 32 Hadamard vectors (instead of 512) for the strange loops. In the text we report about the gain in cost-to-error.}
    \label{fig:noise_reduction}
    \vspace{-0.2cm}
\end{figure}

For certain flavour of quark loops we have also available a calculation made with a different setup. E.g. on the cB211.072.64, for the light disconnected contributions, we have also computed loops without deflation and two stochastic sources, or, for the strange disconnected contributions, we have employed 12 stochastic sources and 32 Hadamard vectors. These different combinations allow us to estimate the improvements due to the noise reduction techniques adopted. We analyse this in Fig.~\ref{fig:noise_reduction} where we compare the error on disconnected contributions for the various windows using the aforementioned statistics for the light- and strange-quark loops. Looking at the light-quark loops, the cost for one stochastic source with deflation is about 35\% cheaper than computing two stochastic sources without deflation, and its error is also 35\% smaller. Summing these up, deflation reduces the cost-to-error ratio by a factor of 3.7$\times$ for this specific quantity and ensemble. Looking at the strange-quark loops, 2 sources with 512 Hadamard vectors require $2.7\times$ more computational resources than 12 sources with 32 Hadamard vector and reduce the errors squared by a factor of $3\times$, resulting in a 11\% reduction of the cost-to-error ratio.

\section{The strange- and charm-quark loops}
\label{sec:disc}

\begin{figure}[htb!]
    \centering
    \includegraphics[width=0.475\linewidth]{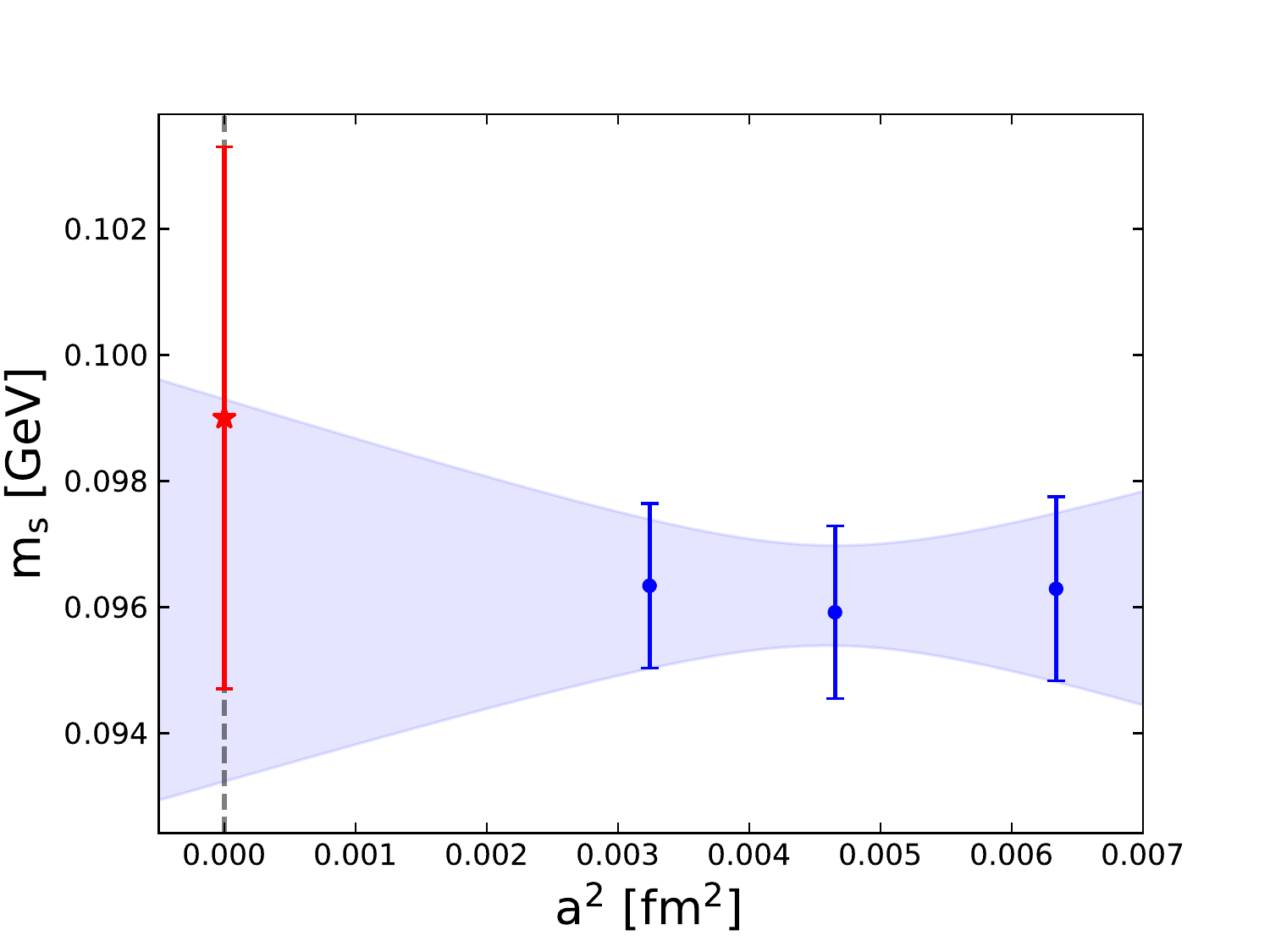}
    \includegraphics[width=0.475\linewidth]{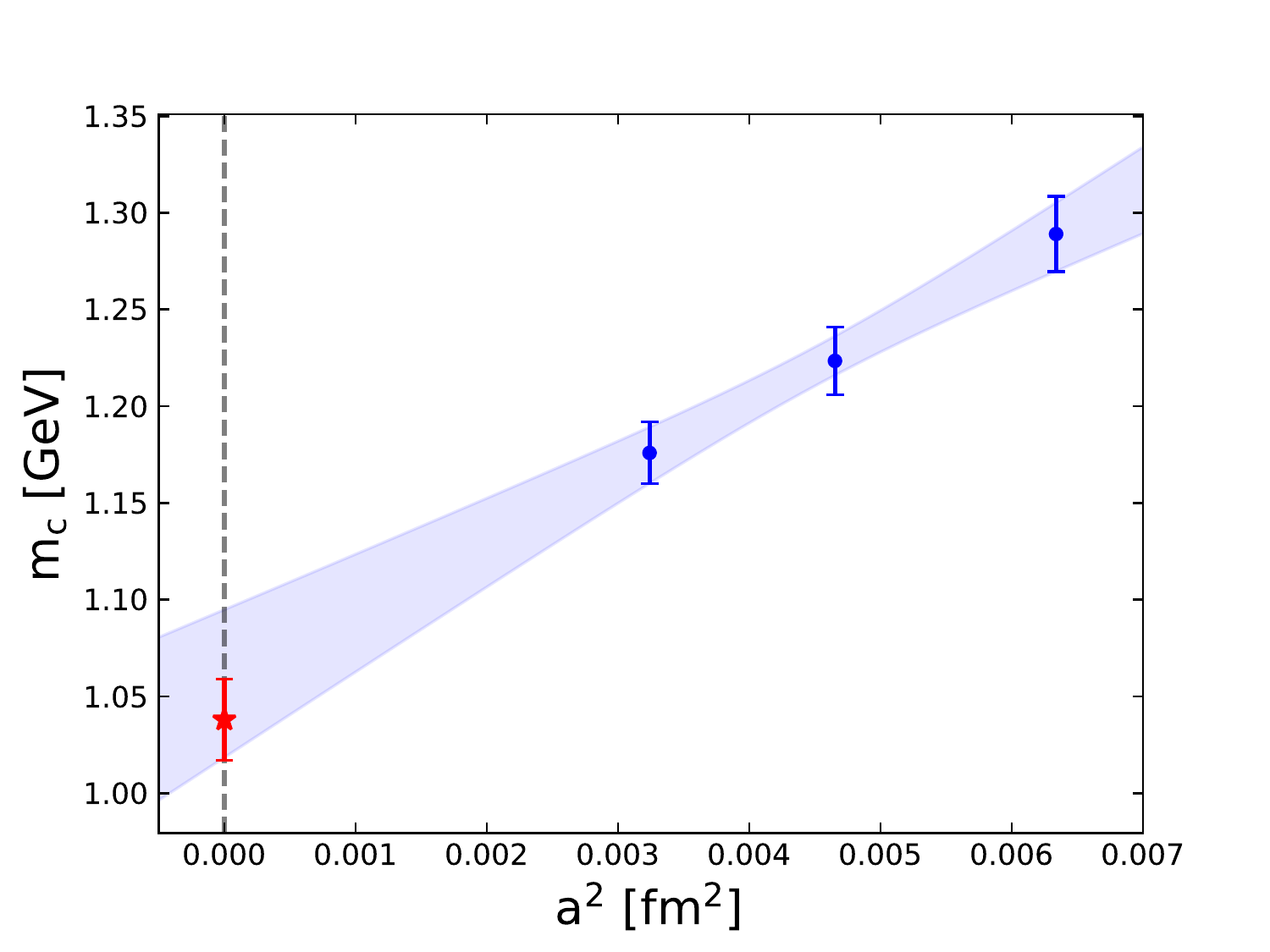}
    \caption{\it \small Renormalized strange (left) and charm (right) quark mass, given respectively in the $\overline{MS}$(2\,GeV) and $\overline{MS}$(3\,GeV) scheme\,\cite{ExtendedTwistedMass:2021gbo,ExtendedTwistedMass_RCs},  tuning the $\Omega$ and $\Lambda_c$ baryon masses, versus the squared lattice spacing. The red stars are the results of the continuum limit extrapolation carried out in Ref.\,\cite{ExtendedTwistedMass:2021gbo}.}
    \label{fig:mu_disconnected}
\end{figure}

The strange- and charm-quark loops are computed at a quark mass obtained by tuning the $\Omega$ and $\Lambda_c$ baryons, respectively, to their physical value. The values of the bare masses for the strange, $a\mu_s$, and for the charm, $a\mu_c$, quarks are listed in Table~\ref{tab:mu_disconnected}. In Fig.~\ref{fig:mu_disconnected}, we show the continuum limit of the renormalized strange and charm quark masses in the $\overline{MS}$(2\,GeV) and $\overline{MS}$(3\,GeV) scheme\,\cite{ExtendedTwistedMass:2021gbo}, respectively. 
 We compare them against the results computed in the continuum limit in Ref.\,\cite{ExtendedTwistedMass:2021gbo}.
 We note that the values of $a\mu_s$ do not show sizable cut-off effects, while $a\mu_c$ does.
 
\begin{table}[htb!]
    \centering
    {\small
    \begin{tabular}{||c||c|c||}
    \hline
        Ensemble & $a \mu_s$ & $a \mu_c$ \\
        \hline
        cB211.072.64 & 0.01860 & 0.249 \\
        cC211.060.80 & 0.01615 & 0.206 \\
        cD211.054.96 & 0.01360 & 0.166 \\
        \hline
    \end{tabular}
    }
    \caption{\it \small Values of the bare quark masses $a\mu_s$ and $a\mu_c$ used for the calculation of strange and charm disconnected contributions.}
    \label{tab:mu_disconnected}
\end{table}

\section{Disconnected contributions}
\label{sec:disconnected}

The quark disconnected contributions to the vector correlator $V(t)$ (see Eq.~(\ref{eq:VV})) is the sum of the six relevant quark disconnected correlators weighted by the appropriate charge factors, namely
\begin{eqnarray}
    V_{\rm disc.}(x) & = & \sum_{i=1,2,3} \Bigg(
    +\frac{1}{9} 
     \left\langle  J_{i}^{\ell\ell}(x) [J_{i}^{\ell\ell}]^\dagger(0) \right \rangle + \frac{1}{9} \left \langle J_{i}^{ss}(x) [J_{i}^{ss}]^\dagger(0) \right \rangle + \frac{4}{9}  \left \langle J_{i}^{cc}(x) [J_{i}^{cc}]^\dagger(0) \right \rangle\\
   & - & \frac{1}{9} \left \langle J_{i}^{\ell \ell}(x) [J_{i}^{ss}]^\dagger(0) + {\rm hc} \right \rangle 
   + \frac{2}{9} \left \langle  J_{i}^{\ell \ell}(x) [J_{i}^{cc}]^\dagger(0) + {\rm hc} \right \rangle
   - \frac{2}{9} \left \langle J_{i}^{ss}(x) [J_{i}^{cc}]^\dagger(0) + {\rm hc} \right \rangle\Bigg)  \, , \nonumber
\label{eq:Vregdetail}     
\end{eqnarray}
where the renormalized vector currents $J_{i}^{ff}$, which are written in terms of Osterwalder-Seiler valence lattice quarks~\cite{Osterwalder:1977pc,Frezzotti:2004wz}, read as
\begin{eqnarray}
    \label{eq:MAVcurr}
    J_{\mu}^{\ell \ell}(x) = Z_V \bar \ell(x) \gamma_\mu  \ell(x)\; , 
    \quad
    J_{\mu}^{ss}(x) = Z_V \bar s(x)\gamma_\mu  s(x) \; , \quad
    J_{\mu}^{cc}(x) = Z_V \bar c(x) \gamma_\mu  c(x) \; . \quad
\end{eqnarray}
For details on the calculation of the renormalization constant $Z_V$ we refer to Ref.~\cite{Alexandrou:2022amy}.
From the vector correlator $V_{disc.}(t)$ the values of $a_\mu^{\rm SD}(disc.)$ and $a_\mu^{\rm W}(disc.)$ are straightforwardly evaluated according to Eq.\,(\ref{eq:amu_w}).
The results for the diagonal and off-diagonal disconnected contributions are summarized in Table~\ref{tab:amuSD_disco} for $a_\mu^{\rm SD}$ and in Table~\ref{tab:amuW_disco} for $a_\mu^{\rm W}$.

\begin{table}[htb!]
    \centering
    \small
    \begin{tabular}{ || c|c|c|c| c| c|c|| } 
    \hline
    Ensemble & $\ell \ell$ & $s s$ & $c c$ & $\ell s$ & $\ell c$ & $s c$ \\
    \hline
    cB211.072.64 & $-3.37~(13)$ & $-2.090~(59)$ & $-1.18~(14)$ & $+5.29~(15)$ & $-1.52~(24)$  & $+1.67~(13)$ \\
    cC211.060.80 & $-3.36~(16)$ & $-2.090~(73)$ & $-0.78~(11)$ & $+5.53~(17)$ & $-1.48~(20)$ & $+1.37~(15)$ \\
    cD211.054.96 & $-3.54~(16)$ & $-2.084~(75)$ & $-0.71~(14)$ & $+5.60~(18)$ & $-1.51~(21)$  & $+1.27~(18)$ \\
    \hline
    \end{tabular}
    \caption{\it \small Summary of the various flavour contributions to $a_{\mu}^{\rm SD}$(disc.) in units of $10^{-12}$ for the cB211.072.64, cC211.060.80 and cD211.054.96 ensembles. The symbols $\ell \ell$, $s s$ and $c c$ denote respectively the flavour-diagonal light, strange and charm contributions, while $\ell s,$ $\ell c$ and $s c$ denote the off-diagonal light-strange, light-charm and strange-charm contributions,  respectively.
	}
    \label{tab:amuSD_disco}
\end{table}
\begin{table}[htb!]
    \centering
    \small
    \begin{tabular}{ || c|c|c|c| c| c|c || } 
    \hline
    Ensemble & $\ell \ell$ & $s s$ & $c c$ & $\ell s$ & $\ell c$ & $s c$  \\
    \hline
    cB211.072.64 & $-1.087~(49)$ & $-0.149~(22)$ & $-0.030~(53)$ & $+0.635~(58)$ & $+0.00~(8)$  & $-0.02~(6)$ \\
    cC211.060.80 & $-1.300~(69)$ & $-0.159~(27)$ & $-0.033~(49)$ & $+0.726~(81)$ & $-0.03~(7)$ & $+0.04~(7)$  \\
    cD211.054.96 & $-1.201~(73)$ & $-0.149~(29)$ & $+0.018~(54)$ & $+0.627~(81)$ & $+0.02~(8)$  & $-0.02~(7)$ \\
    \hline
    \end{tabular}
    \caption{\it \small The same as in Table\,\ref{tab:amuSD_disco}, but for the various flavour contributions to $a_\mu^{\rm W}$(disc.) in units of  $10^{-10}$.}
    \label{tab:amuW_disco}
\end{table}

\begin{table}[htb!]
    \centering
    \small
    \begin{tabular}{ ||c|c|c|| } 
    \hline
    Ensemble & $a_\mu^{\rm SD}$(disc.) & $a_\mu^{\rm W}$(disc.)  \\
    \hline
    cB211.072.64 & $-1.20~(23)\cdot 10^{-12}$ & $-0.651~(93)\cdot 10^{-10}$ \\
    cC211.060.80 & $-0.80~(18)\cdot 10^{-12}$ & $-0.762~(75)\cdot 10^{-10}$  \\
    cD211.054.96 & $-0.96~(20)\cdot 10^{-12}$ & $-0.701~(80)\cdot 10^{-10}$ \\
    \hline
    Continuum &  $- 0.6~(5) \cdot 10^{-12}$ & $- 0.78~(21) \cdot 10^{-10}$ \\
    \hline
    \end{tabular}
    \caption{\it \small Final results for the disconnected contributions to the short and intermediate distance windows}
    \label{tab:amu_disco_final}
\end{table}

\begin{figure}[htb!]
\begin{center}
\includegraphics[width=0.75\linewidth]{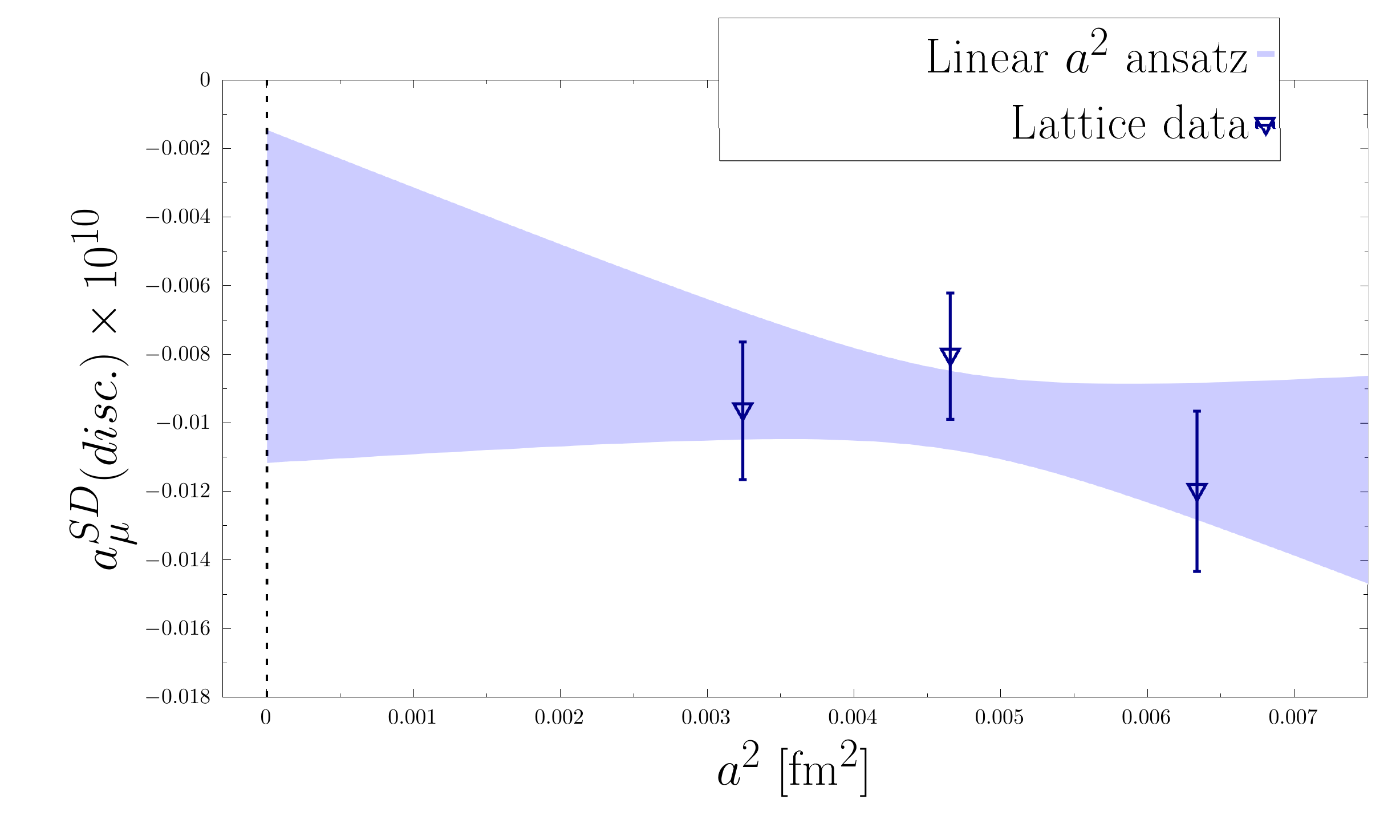}
\includegraphics[width=0.75\linewidth]{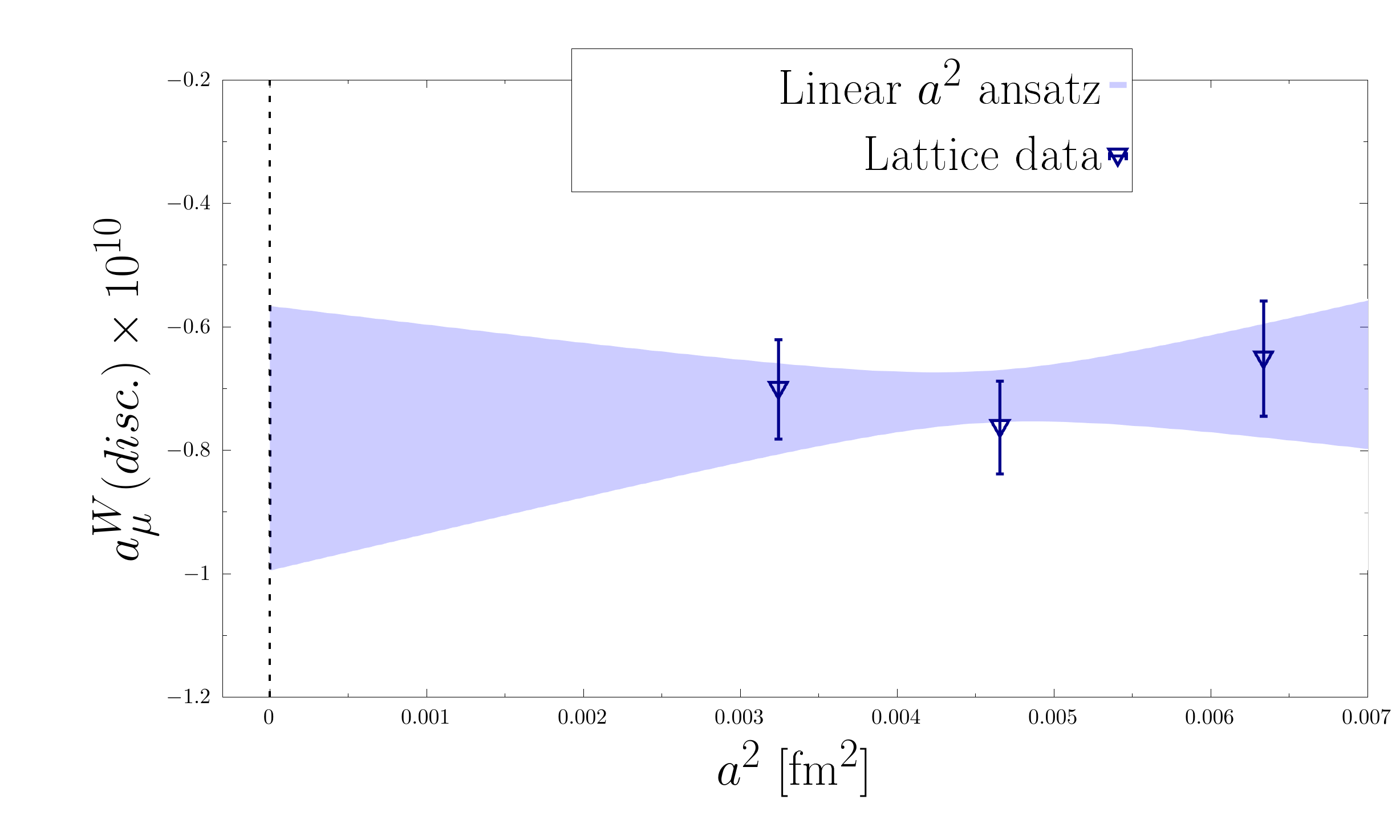}
\caption{\it \small Top panel: The quark-loop disconnected contribution to the short time-distance window, $a_\mu^{\rm SD}$, versus the squared lattice spacing $a^2$ in physical units. Bottom panel: the same as in the top panel, but for the intermediate window $a_\mu^{\rm W}$. The blue band corresponds to the extrapolation performed using a linear fit Ansatz in $a^2$.}
\label{fig:SDW_disco_cont_lim}
\end{center}
\end{figure}

In Table~\ref{tab:amu_disco_final} we give the sum of all contributions and the result of the continuum limit that we take as final value. In Fig.~\ref{fig:SDW_disco_cont_lim}, we show the continuum limit extrapolation for the disconnected contributions to $a_\mu^{\rm SD}$ and $a_\mu^{\rm W}$. Qualitatively, for $a_\mu^{\rm W}$ the light-light contribute +150\% of the total disconnected contribution, the strange-light -80\% and the strange-strange +30\%. All other combinations are consistent with zero within the errors. We do not observe sizable cutoff effects at this level of precision.
The disconnected contribution to $a_\mu^{\rm SD}$ is very small, being approximately forty times smaller as compared to our error on the light-connected contribution to $a_\mu^{\rm SD}$. Given the available data, we  perform only two continuum extrapolation, using either a constant fit Ansatz or a linear one in $a^2$. The latter fit Ansatz is adopted in view of the expected automatic O(a) improvement~\cite{Frezzotti:2004wz} of the relevant two-point
correlation functions of the renormalized vector currents in Eq.~\eqref{eq:MAVcurr}. The extrapolated values using these two fitting procedures are in agreement with each other. The one obtained from the linear fit in $a^2$ has a larger statistical uncertainty and conservatively  we take it as our final estimate.

\begin{table}[htb!]
\centering
\small
    \begin{tabular}{||c|c||}
    \hline
    ~ Ref. ~ & $a_\mu^{\rm W}(disc.)$  \\
  \hline \hline
  this work & $~-0.78~(0.21)\cdot 10^{-10}$ \\
  \hline
 BMW~\mbox{\cite{Borsanyi:2020mff}} & $~-0.85~(0.06)\cdot 10^{-10}$ \\
  \hline
 ~CLS/Mainz~\mbox{\cite{Ce:2022kxy}} & $~-0.81~(0.09)\cdot 10^{-10}$ \\
  \hline \hline
  average & $~-0.83~(0.05)\cdot 10^{-10}$ \\
  \hline
    \end{tabular}
\caption{\it \small Disconnected contribution to the intermediate time-distance window $a_\mu^{\rm W}$ obtained in this work and in Refs.\,\cite{Borsanyi:2020mff, Ce:2022kxy}. The last row lists the average of all lattice results made following the PDG approach.}
\label{tab:comparison_LQCD}
\end{table} 

\section{Conclusions}

In Table~\ref{tab:comparison_LQCD} we compare our results with those obtained by BMW~\cite{Borsanyi:2020mff} and CLS/Mainz~\cite{Ce:2022kxy} on the disconnected contribution to the intermediate window. The largest systematic effect on our error comes from the continuum extrapolation. We observe a remarkable agreement among all available lattice results. In the future we plan to extended this calculation with a further lattice spacing in order to improve our continuum extrapolation.

\section*{Acknowledgments}

We thank all members of ETMC for the most enjoyable collaboration. 
We thank the developers of the QUDA~\cite{Clark:2009wm, Babich:2011np, Clark:2016rdz} library for their continued support, without which the calculations for this project would not have been possible.
S.B.~and J.F.~are supported by the H2020 project PRACE 6-IP (grant agreement No.~82376) and the EuroCC project (grant agreement No.~951740). We acknowledge support by the European Joint Doctorate program STIMULATE grant agreement No.~765048. P.D. acknowledges support from the European Unions Horizon 2020 research and innovation programme under the Marie Sk\l{}odowska-Curie grant agreement No.~813942 (EuroPLEx) and also support from INFN under the research project INFN-QCDLAT.
K.H.~is co-funded by the European regional development fund and the Republic of Cyprus through the Research and Innovation Foundation under contract number POST-DOC/0718/0100, under contract number CULTURE-AWARD-YR/0220/0012 and by the EuroCC project (grant agreement No.~951740). 
R.F.~acknowledges partial support from the University of Tor Vergata program “Beyond Borders/ Strong Interactions: from Lattice QCD to Strings, Branes and Holography".
F.S., G.G.~and S.S.~are supported by the Italian Ministry of University and Research (MIUR) under grant PRIN20172LNEEZ. 
F.S.~and G.G.~are supported by INFN under GRANT73/CALAT.
This work is supported by the Deutsche Forschungsgemeinschaft (DFG, German Research Foundation) and the NSFC through the funds provided to the Sino-German Collaborative Research Center CRC 110 “Symmetries and the Emergence of Structure in QCD” (DFG Project-ID 196253076 - TRR 110, NSFC Grant No.~12070131001).
The authors gratefully acknowledge the Gauss Centre for Supercomputing e.V.~(www.gauss-centre.eu) for funding the project pr74yo by providing computing time on the GCS Supercomputer SuperMUC at Leibniz Supercomputing Centre (www.lrz.de), as well as computing time projects 
on the GCS supercomputers JUWELS Cluster and JUWELS Booster~\cite{JUWELS} at the J\"ulich Supercomputing Centre (JSC) and time granted by the John von Neumann Institute for Computing (NIC) on the supercomputers JURECA and JURECA Booster~\cite{Jureca}, also at JSC. Part of the results were created within the EA program of JUWELS Booster also with the help of the JUWELS Booster Project Team (JSC, Atos, ParTec, NVIDIA). We further acknowledge computing time granted on Piz Daint at Centro Svizzero di Calcolo Scientifico (CSCS) via the project with id s702. The authors acknowledge the Texas Advanced Computing Center (TACC) at The University of Texas at Austin for providing HPC resources that have contributed to the research results. The authors gratefully acknowledge PRACE for awarding access to HAWK at HLRS within the project with Id Acid 4886.

{
\bibliography{biblio}
\bibliographystyle{JHEP}
}

\end{document}